\documentclass[12pt]{article}
\usepackage[english]{babel}
\usepackage{isolatin1}
\usepackage{amsmath}
\usepackage{epsfig}
\usepackage{tabularx} 
\usepackage{changebar}

\begin{document}
\title{\large A Mean Field Theory approach to a Dipole-Quadrupole 
interaction model}
\author{R.P.B. dos Santos
\footnote{E-mail: paupitz@if.usp.br}\\
	{\small\sl Instituto de Física da Universidade de S\~ao Paulo}\\
	{\small\sl Caixa Postal 66318, 05315-970 S\~ao Paulo, SP, Brazil}
}
\maketitle \normalsize \abstract{We study a model initially proposed
to describe a mixture of $\mathrm{(CO)}_{1-x}\mathrm{(N_2)}_x$
adsorbed on exfoliated graphite. The approach used here is that of
mean field theory. The Mean Field equations and the Helmholtz Free
Energy are found. Phase diagrams are calculated too, and it
is possible to find an analytic expression for the second order phase
transition line.  }

\section{Introduction}

To calculate the fundamental equation of a many particles system, it is
necessary first to determinate the allowed energy levels and make the
sum of the partition function.  Except for a small class of systems,
that sum cannot be calculated exactly. One of the solutions for that
problem is to look for approximated solutions, in our case we use the
so called Mean Field Theory. The importance of this kind of
calculation is that it gives us qualitative information about the
critical behaviour of the system.

Here we show results for a model of solid mixtures of molecules. We
assume that one of the molecules have dipole and quadrupole moments
($\mathrm{CO}$ molecules) while the other one has only a quadrupole
moment ($\mathrm{N_2}$). This simple model is in good agreement with
experimental results \cite{pereyra:1993}. The interactions between the
$\mathrm{CO}$ molecules are assumed to be antiferroelectric, but
between these and the $\mathrm{N_2}$ molecules it is proposed a pseudo
dipole-quadrupole coupling. The $\mathrm{N_2}$ molecules act like the
random fields of the Random Field Ising Model. It is supposed that a
crystal of polar molecules diluted with molecules which have
quadrupole moments, can have a behaviour qualitatively similar to that
predicted in random field models \cite{wiechert:1993}.

There are Monte Carlo results for this model
\cite{pereyra:1993,pereyra:1995} which agree very well with
experiments. Despite the fact that there are simulations and real
experiments for this model, we don't have analytical results, exact or
not, for this model yet. In order to improve our understanding about
this kind of system, it is interesting to search for such results. The
simplest way to find approximated results is to calculate a Mean Field
Approximation for the model, what we do in the next sections.  The
paper is organized as follows. In section 2 we present the
Dipole-Quadrupole model. In section $3$ we calculate the mean field
approximation using the {\sl Bogoliubov Variational Theorem}
\cite{callen:1985} obtaining the mean field equations and the
Helmholtz Free Energy of the system. In section $4$, we find analytic
expressions for the transition lines and these lines are showed for
several values of model parameters.

\section{Dipole-Quadrupole model}

This model was proposed to describe a mixture of
$\mathrm{(CO)}_{1-x}\mathrm{(N_2)}_x$ adsorbed on exfoliated graphite
\cite{pereyra:1993}. We associate for each site of the lattice a spin
$S_i=-1,0,1$. $S_i=\pm 1$ represents the
electrical dipole orientations if the site is occupied by a
$\mathrm{CO}$ molecule and $S_i=0$ if the site is occupied by a
$\mathrm{N_2}$ molecule. The interaction between the molecules of
$\mathrm{CO}$ is antiferromagnetic and the interaction for
$\mathrm{CO}$ - $\mathrm{N_2}$ pairs is described like a pseudo
dipole-quadrupole one \cite{pereyra:1993,pereyra:1995}.  The proposed
Hamiltonian is
\begin{equation}
\mathcal{H}=-J\sum_{\langle ij\rangle}S_iS_j-J_1\sum_{\langle ij\rangle}
S_i(1-S_{j}^2),
\end{equation}
where $J<0$ represents the antiferromagnetic coupling, $J_1$ describes
the dipole-quadrupole interaction and $S_i=0,\pm 1$. The symbol
$\langle \cdots\rangle$ denotes first neighbours summation. We rewrite
the Hamiltonian in the form
\begin{equation}
\mathcal{H}=-J\sum_{\langle ij\rangle}S_iS_jc_ic_j-J_1\sum_{\langle ij\rangle}
S_ic_i(1-c_j),
\end{equation}
where $S_i=\pm 1$ and $c_i=1$ if the site is occupied by a
$\mathrm{CO}$ molecule and $c_i=0$ otherwise. The probability
distribution for the occupation variables is
\begin{equation}
P(c_i)=p\delta (c_i-1)+(1-p)\delta (c_i),
\end{equation}
with $0\leq p\leq 1$.  The presence of a antiferromagnetic interaction
suggests that the system can spontaneously be subdivided in two
subsystems, represented by two sub-lattices (figure 1). 
\begin{figure}[h]
\begin{center}
\epsfig{file=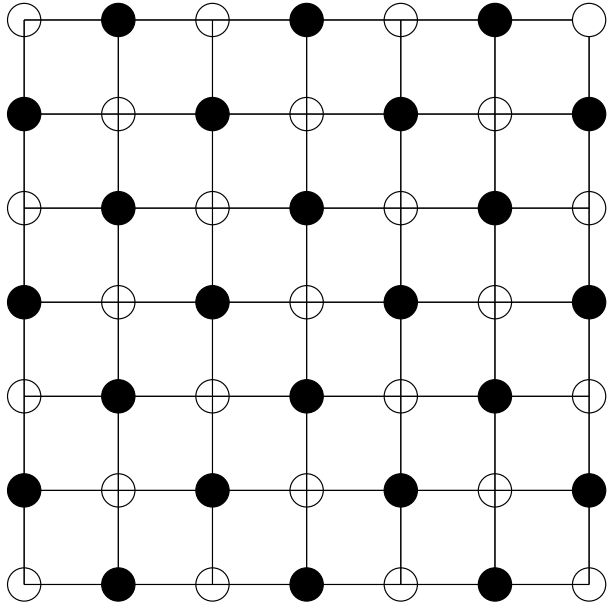,height=8cm,width=8cm,angle=270}
\end{center}
\caption{Sub-lattices A and B in a bidimensional lattice.}
\end{figure}

\section{Mean field calculations}

We propose an unperturbed Hamiltonian $H_0$ of the form
\begin{equation}
H_0=-(J_1+\eta_A)\sum_{i\in A}S_ic_i-(J_1+\eta_B)\sum_{i\in B}S_ic_i,
\end{equation}
where $\eta_A$ and $\eta_B$ are variational parameters and A and B
indicate the inter-penetrating sub-lattices, each one with $N/2$
sites. We choose $H_0$ to be, at the same time, soluble and similar
in some manner to the original Hamiltonian. 

The Bogoliubov Inequality reads
\begin{equation}
\big\langle F\big\rangle \leq \big\langle F_0\big\rangle +\Big\langle\langle H-H_0\rangle_t\Big\rangle,
\end{equation} 
where $F_0$ is the Helmholtz Free Energy calculated for the
unperturbed system and $F$ indicates the complete Helmholtz Free
Energy. The symbol $\langle\cdots\rangle$ means a configurational
average, while $\langle\cdots\rangle_t$ indicates the thermal average
weighted by the factors $\exp{(-\beta E_i)}$ where $E_i$ is the
eigenvalue of energy with indice $i$. The partition function of the
soluble Hamiltonian is
\begin{equation}
\begin{split}
Z_0=\mathrm{Tr}\big\{e^{-\beta H_0}\big\}
=&\prod_{i\in A}\prod_{j\in B}\biggl[ 2\cosh{\biggl( \beta c_i(J_1+\eta_A)
\biggr)}\biggr] \\ 
&\times\biggl[ 2\cosh{\biggl( \beta c_j(J_1+\eta_B)\biggr) }\biggr],
\end{split}
\end{equation}
and the mean Free Energy calculated with that partition function becomes
\begin{equation}
\begin{split}
\langle F_0\rangle=-\frac{N}{2\beta}\Biggl[& \Big< \ln{\Big(2\cosh{\beta c_i(J_1+\eta_A)}\Big)}\Big>\\
&+ \Big< \ln{\Big(2\cosh{\beta c_i(J_1+\eta_B)}\Big)}\Big>
\Biggr],
\end{split}
\end{equation}
the second part of inequality (5) is written  
\begin{equation}
\begin{split} 
\Big<\langle H-H_0\rangle_t\Big> = &\Big<
J\sum_{i\in A}\sum_{j\in B} \langle
S_iS_jc_ic_j\rangle_t+J_1\sum_{i\in A}\sum_{j\in B}\langle S_ic_ic_j
\rangle_t+ \\
&\eta_A\sum_{i\in A}\langle S_ic_i\rangle_t+\eta_B\sum_{j\in
B}\langle S_jc_j\rangle_t 
\Big> .
\end{split} 
\end{equation}
If we adjust the variational parameters $\eta_A$ and $\eta_B$ using the
stationary condition for the Free Energy and define the magnetizations 
\begin{equation}
\begin{split} 
m_A = \Biggl< c_i \tanh{\biggl[ c_i\beta(H+\eta_A) \biggr]}\Biggr>,
\end{split} 
\end{equation}
and 
\begin{equation}
\begin{split} 
m_B = \Biggl< c_i \tanh{\biggl[ c_i\beta(H+\eta_B) \biggr]}\Biggr>,
\end{split} 
\end{equation}
it is possible to write the Bogoliubov Inequality like
\begin{equation}
\begin{split} 
\bigl< F \bigr> \leq & -\frac{N}{2\beta}\biggl[ \Bigl<
\ln{\Bigl(2\cosh{\bigl(\beta c_i(J_1+\eta_A)\bigr)}\Bigr)}\Bigr>+ \Bigl<\ln{\Bigl(2\cosh{\bigl(\beta c_i(J_1+\eta_B)\bigr)}\Bigr)}\Bigr>
\Biggr] \\
&+J\frac{Nz}{2}m_Am_B+J_1\frac{Nz}{4}\Bigl[ \langle c_i\rangle m_A
+\langle c_i\rangle m_B\Bigr]+\eta_A\frac{N}{2}m_A+\eta_B\frac{N}{2}m_B.
\end{split} 
\end{equation}
This expression must be minimized in order to estimate the Free
Energy. Differentiating the equation and imposing the stationary
condition we find the variational parameters
\begin{equation}
\begin{split} 
\eta_A=-Jzm_B-J_1\frac{z}{2}\langle c_i\rangle
\end{split} 
\end{equation}
and 
\begin{equation}
\begin{split} 
\eta_B=-Jzm_A-J_1\frac{z}{2}\langle c_i\rangle
\end{split} 
\end{equation}
then
\begin{equation}
\begin{split} 
\langle F \rangle = &-\frac{1}{2\beta}\biggl[ \Bigl<
\ln{\Bigl(2\cosh{\bigl(\beta c_i(J_1-\frac{J_1z}{2}\langle c_i\rangle-Jzm_B)
\bigr)}\Bigr)}\Bigr>\\
&\quad\quad+ 
\Bigl<\ln{\Bigl(2\cosh{\bigl(\beta c_i(J_1-\frac{J_1z}{2}\langle c_i\rangle-
Jzm_A)\bigr)}\Bigr)}\Bigr>
\Biggr] \\
&-\frac{1}{2}Jzm_Am_B. 
\end{split} 
\end{equation}
Again, the antiferromagnetic interaction suggests we should define
\begin{equation}
\begin{split} 
M=\frac{m_A+m_B}{2},
\end{split} 
\end{equation}
where $M$ is the total magnetization and 
\begin{equation}
\begin{split} 
m_s=\frac{m_A-m_B}{2},
\end{split} 
\end{equation}
It is also convenient to define
\begin{equation}
\begin{split} 
t=\frac{1}{\beta |J|z},
\end{split} 
\end{equation}
and  
\begin{equation}
\begin{split} 
h=\frac{J_1}{|J|z}\Bigl( 1-\frac{z}{2}\langle c_i\rangle\Bigr),
\end{split} 
\end{equation}
we find 
\begin{equation}
\begin{split} 
\langle F \rangle = &-\frac{1}{2\beta}\biggl[ \Bigl<
\ln{\Bigl(2\cosh{\bigl(\frac{c_i}{t}(h-M+m_s)
\bigr)}\Bigr)}\Bigr>\\
&\quad\quad+ 
\Bigl<\ln{\Bigl(2\cosh{\bigl( \frac{c_i}{t}(h-M-m_s)\bigr)}\Bigr)}\Bigr>
\Biggr] \\
&-\frac{1}{2\beta t}M^2+\frac{1}{2\beta t}m_s^2, 
\end{split} 
\end{equation}
(15) and (16) define the mean field equations for this model 
\begin{equation}
\begin{split} 
M = \frac{1}{2}\biggl[ \Bigl<c_i\tanh{\Bigl(\frac{c_i}{t}(h-M+m_s)\Bigr)}
\Bigr> + \Bigl<c_i\tanh{\Bigl(\frac{c_i}{t}(h-M-m_s)\Bigr)}\Bigr>
\Biggr],
\end{split} 
\end{equation}
and 
\begin{equation}
\begin{split} 
m_s = \frac{1}{2}\biggl[ \Bigl<c_i\tanh{\Bigl(\frac{c_i}{t}(h-M+m_s)\Bigr)}
\Bigr> - \Bigl<c_i\tanh{\Bigl(\frac{c_i}{t}(h-M-m_s)\Bigr)}\Bigr>
\Biggr].
\end{split} 
\end{equation}
These coupled equations can be solved for $M$ and $m_s$ using a
iterative scheme. 

\section{Critical lines }
The existence of a critical line, critical points and the
determination of transition order are made by means of a Landau
expansion \cite{galam:1998}. Then we write the magnetization in the
form
\begin{equation}
\begin{split} 
M=M_0+m,
\end{split} 
\end{equation}
where $M_0$ is a paramagnetic solution which is a solution of 
\begin{equation}
\begin{split} 
M_0=\Bigl<c_i\tanh{\Bigl(\frac{c_i}{t}(h-M_0)\Bigr)}\Bigr>.
\end{split} 
\end{equation}
We define the quantities $m_1=m-m_s$ and $m_2=m+m_s$ to expand the
right hand side of equation (22) near a paramagnetic solution. Then
\begin{equation}
\begin{split} 
M=\sum_{n=0}^{\infty}A_n(m_1^n+m_2^n),
\end{split} 
\end{equation}
eliminating the term corresponding to $n=0$ we find
\begin{equation}
\begin{split} 
m=\sum_{n=1}^{\infty}A_n(m_1^n+m_2^n),
\end{split} 
\end{equation}
it is useful to expand the {\sl staggered magnetization} too
\begin{equation}
\begin{split} 
m_s=\sum_{n=1}^{\infty}A_n(m_1^n-m_2^n).
\end{split} 
\end{equation}
We determine the coefficients $A_n$ in these expressions
\begin{equation}
\begin{split} 
A_1=-\frac{1}{t}\biggl< c_i\Bigl(1-T_2\Bigr)\biggr>,
\end{split} 
\end{equation}
\begin{equation}
\begin{split} 
A_2=-\frac{1}{2t^2}\biggl< c_i^2\Bigl(T_1-T_3\Bigr)\biggr>,
\end{split} 
\end{equation}
\begin{equation}
\begin{split} 
A_3=\frac{1}{3t^3}\biggl< c_i^3\Bigl(1-4T_2+3T_4\Bigr)\biggr>,
\end{split} 
\end{equation}
\begin{equation}
\begin{split} 
A_4=\frac{1}{3t^4}\biggl< c_i^4\Bigl(3T_5-5T_3+2T_1\Bigr)\biggr>,
\end{split} 
\end{equation}
\begin{equation}
\begin{split} 
A_5=\frac{1}{15t^5}\biggl< c_i^5\Bigl(15T_6-30T_4+17T_2-2\Bigr)\biggr>,
\end{split} 
\end{equation}
where
\begin{equation}
\begin{split} 
T_k=\tanh^k{\Bigl[\frac{1}{t}(h-M_0)\Bigr]}.
\end{split} 
\end{equation}
It is possible to calculate $m$ in function of $m_s$
\begin{equation}
\begin{split} 
m=B_1m_s^2+B_2m_s^4+B_3m_s^6+\cdots,
\end{split} 
\end{equation}
if we substitute this expansion in equation (25) and equal terms of
same power in $m_s$ we find the expressions for the coefficients $B_i$
in terms of $A_i$
\begin{equation}
\begin{split} 
B_1=\frac{A_2}{1-A_1},
\end{split} 
\end{equation}
\begin{equation}
\begin{split} 
B_2=\frac{1}{1-A_1}\Biggl( \frac{A_2^3}{(1-A_1)^2}+A_4+3
\frac{A_2A_3}{1-A_1}\Biggr),
\end{split} 
\end{equation}
\begin{equation}
\begin{split} 
B_3=\frac{1}{1-A_1}&\Biggl[ \frac{2A_2^2}{(1-A_1)^2}\Biggl( 
\frac{A_2^3}{(1-A_1)^2}+A_4+3\frac{A_2A_3}{1-A_1}\Biggr)\\
&+
\frac{A_2^3A_3}{(1-A_1)^3}+\frac{6A_2^4A_4}{(1-A_1)^4}\frac{5A_2A_5}{1-A_1}
\Biggr].
\end{split} 
\end{equation}
Here, substituting (33) in (26) we arrive at expression
\begin{equation}
\begin{split} 
am_s+bm_s^3+cm_s^5+\cdots =0,
\end{split} 
\end{equation}
and finally at the coefficients
\begin{equation}
\begin{split} 
a=-(1+A_1),
\end{split} 
\end{equation}
\begin{equation}
\begin{split} 
b=-\Biggl( 2\frac{A_2^2}{1-A_1}+A_3\Biggr),
\end{split} 
\end{equation}
\begin{equation}
\begin{split} 
c=-\Biggl[ 2\frac{A_2}{1-A_1}
\Biggl( \frac{A_2^3}{(1-A_1)^2}+A_4+3\frac{A_2A_3}{1-A_1}\Biggr)
+3\frac{A_2^2A_3}{(1-A_1)^2}+4\frac{A_2A_4}{1-A_1}+A_5\Biggr].
\end{split} 
\end{equation}
A Second order transition is determined by the condition $a=0$ and
$b>0$. The existence of a tricritical point will be determined by
$a=b=0$ with $c>0$ \cite{kincaid:1975}.

With these conditions found, we calculate configurational mean values
as indicated for the distribution eq. (3) to find that the critical
line condition is equivalent to
\begin{equation}
\begin{split} 
1-\frac{p}{t}\biggl[ 1-\tanh^2{\Bigl( \frac{h-M_0}{t}\Bigr)}\biggr]=0,
\end{split} 
\end{equation}
which, together with the configurational mean value of expression (23) gives us
\begin{equation}
\begin{split} 
\frac{M_0}{p}=\Bigl[ 1-\frac{t}{p}\Bigr]^{1/2},
\end{split} 
\end{equation}
from which we note a limitation on reduced temperature represented by 
$t\leq p$. Now the critical line equation is 
\begin{equation}
\begin{split} 
h=t\tanh^{-1}{\Bigl[ 1-\frac{t}{p}\Bigr]^{1/2}}+p\Bigl[ 1-
\frac{t}{p}\Bigr]^{1/2},
\end{split} 
\end{equation}
and, finally
\begin{equation}
\begin{split} 
\frac{J_1}{|J|}=\frac{z}{\big(1-\frac{zp}{2}\big)}\Bigg[t\tanh^{-1}{\Big( 1-
\frac{t}{p}\Big)^{1/2}}+p\Big( 1-\frac{t}{p}\Big)^{1/2}\Bigg].
\end{split} 
\end{equation}
We should note here that, for a particular lattice of coordination
number $z$, there exist a critical value for the occupation $p$
where there is a divergence. This point is located at 
\begin{equation}
p_c = \frac{2}{z}.
\end{equation}

The condition $c>0$ for existence of a tricritical point
\begin{equation}
\begin{split} 
\frac{2}{3t}\cosh^4{\Bigl(\frac{h-M_0}{t}\Bigr)}+\Bigl( \frac{2p}{3t^2}-
\frac{p}{2}-\frac{1}{t}\Bigr)\cosh^2{\Bigl( \frac{h-M_0}{t}\Bigr)}+
\frac{p}{2}-\frac{p}{t^2}>0.
\end{split} 
\end{equation}
Is not satisfied for this model.
\begin{figure}
\begin{center}
\epsfig{file=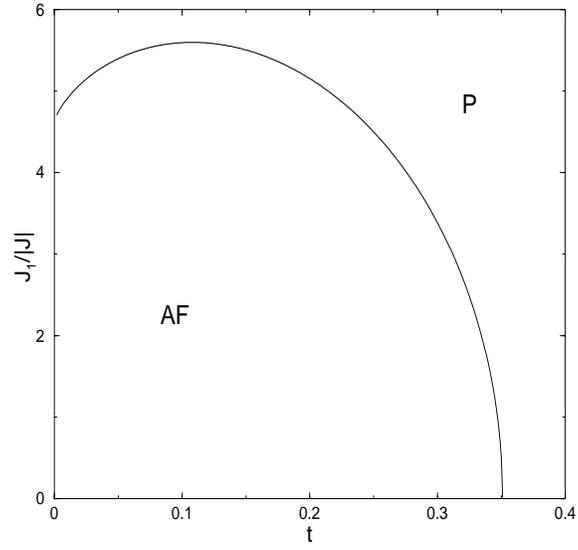,height=7.5cm,width=7.5cm,angle=270}
\end{center}
\caption{Second order critical line for $z=4$ and $p=0.35$.}
\end{figure}

\begin{figure}
\begin{center}
\epsfig{file=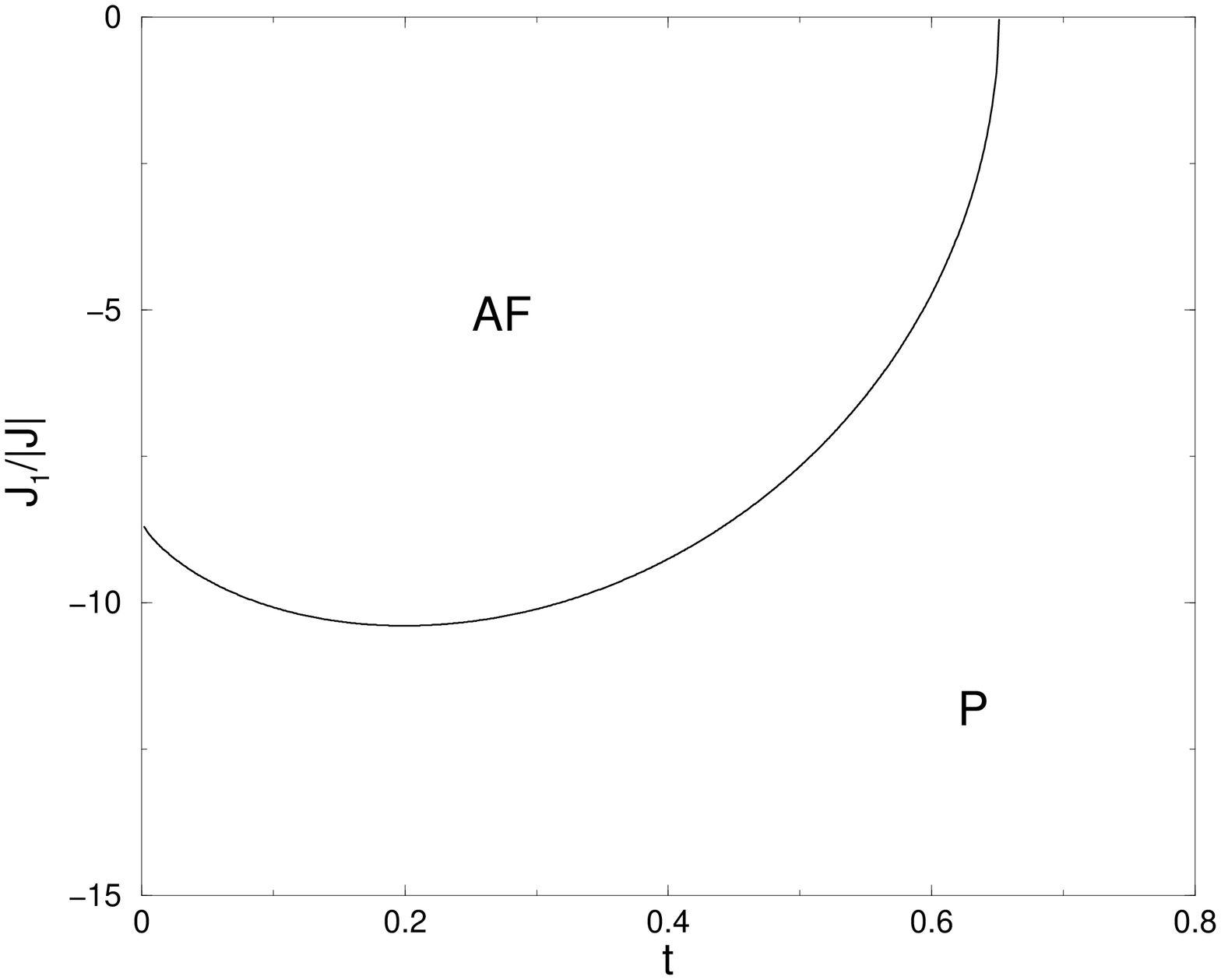,height=7.5cm,width=7.5cm,angle=270}
\end{center}
\caption{Second order critical line for $z=4$ and $p=0.65$.}
\end{figure}

\begin{figure}
\begin{center}
\epsfig{file=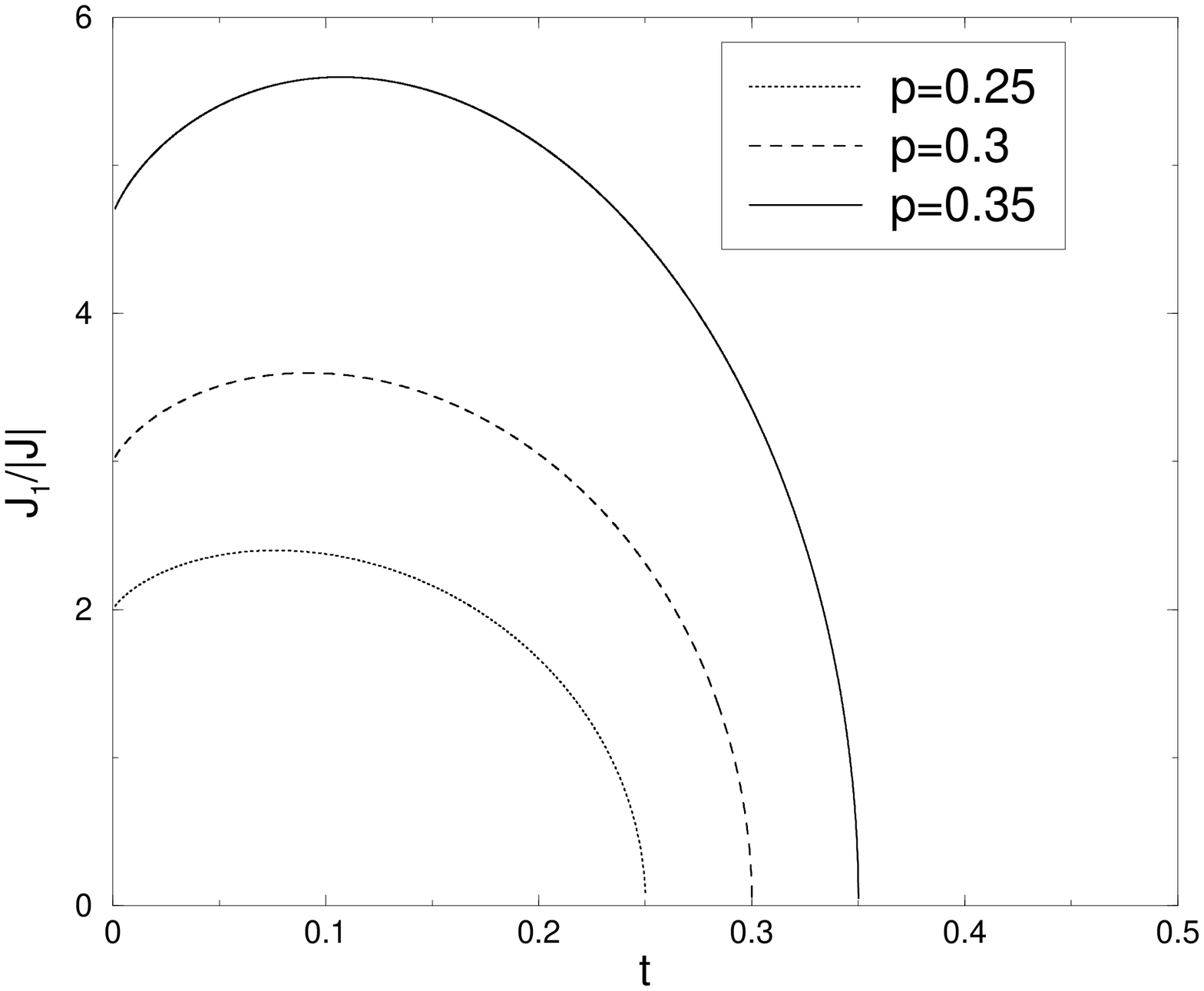,height=7.5cm,width=7.5cm,angle=270}
\end{center}
\caption{Comparison for three values of the occupation rate which are
less than the critical occupation.}
\end{figure}

\begin{figure}
\begin{center}
\epsfig{file=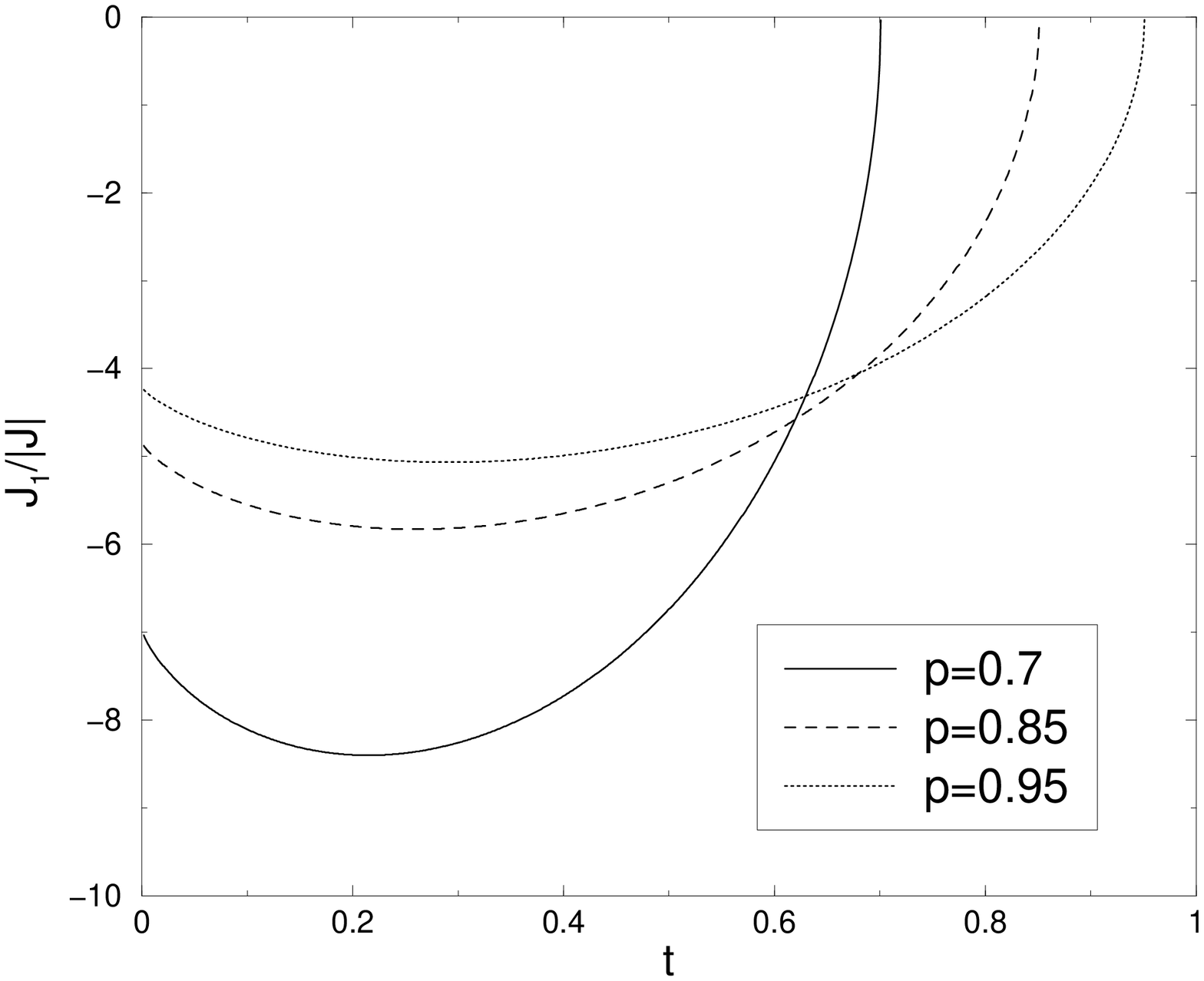,height=7.5cm,width=7.5cm,angle=270}
\end{center}
\caption{Comparison for three values of the occupation rate which are
bigger than the critical occupation.}
\end{figure}
In figures (2) and (3) we have two phase diagrams (one for
$p<2/z$ and one for $p>2/z$) for the model in the case
$z=4$ indicating two possible phases: Antiferromagnetic (AF) and
Paramagnetic (P).
In figures (4) and (5) we compare the phase
diagrams for different values of occupation $p$. 
\vfill
\null

\section{Conclusions}
Preliminary mean field results for the model Hamiltonian (2) was
presented in this paper. This model describes dipolar systems diluted
with quadrupolar molecules. It was possible to determine, using the
Bogoliubov Inequality, the mean field equations for the model and the
Helmholtz Free Energy. The last one was calculated both, in terms of the
magnetizations of the sub-lattices $m_a$ and $m_b$ and in terms of
total magnetization $M$ and {\sl staggered magnetization} $m_s$. We
have also calculated the analytical form for the second order
$\frac{J_1}{|J|}\times t$ critical line. We found a critical value of
occupation probability $p_c$ in which there is a divergence in
expression (44).  The critical lines are remarkably different in the
regimes of $p<p_c$ and $p>p_c$. Another interesting fact is that this
model does not have any tricritical or multicritical points, at least
in the mean field approximation. 

We hope this study is able to stimulate theoretical physicists to
search for more theoretical results about this kind of system. It is
specially interesting to learn about the possible connections between
these and random field models.
 
\section*{Acknowledgments}
We would like to thanks {\bf FAPESP} (Fundaç\~ao de Amparo à Pesquisa
do Estado de S\~ao Paulo) for financial support during this research.

\end{document}